\documentclass[useAMS,usenatbib]{mn2e}
\usepackage{graphicx}                                            
\usepackage{times}
\usepackage{epsfig}
\usepackage{rotating}

\def\lesssim{\mathrel{\hbox{\rlap{\hbox{\lower4pt\hbox{$\sim$}}}\hbox{$<$}}}}
\let\la=\lesssim
\def\gtrsim{\mathrel{\hbox{\rlap{\hbox{\lower4pt\hbox{$\sim$}}}\hbox{$>$}}}}
\let\ga=\gtrsim

\def\sol{~\mathrm{M}_\odot}

\title[The rms-intensity diagram]{Fast variability as a tracer of accretion regimes in black hole transients}
\author[Mu\~noz-Darias, Motta \& Belloni]{T.~Mu\~noz-Darias$^{1}$ \thanks{E-mail: tmd@brera.inaf.it}, S.~Motta$^{1,2}$ and T.~M.~Belloni$^{1}$\\
$^{1}$INAF-Osservatorio Astronomico di Brera, Via E. Bianchi 46, I-23807 Merate (LC), Italy\\
$^{2}$Universit\`a dell'Insubria, Via Valleggio 11, I-22100 Como, Italy}

\begin{document}
\maketitle 
\begin{abstract}
We present the rms-intensity diagram for black hole transients. Using observations taken with the Rossi X-ray timing explorer we study the relation between the root mean square (rms) amplitude of the variability and the net count-rate during the 2002, 2004 and 2007 outbursts of the black hole X-ray binary GX 339-4. We find that the rms-flux relation previously observed during the hard state in X-ray binaries does not hold for the other states, when different relations apply. These relations can be used as a good tracer of the different accretion regimes. We identify the hard, soft and intermediate states in the rms-intensity diagram. Transitions between the different states are seen to produce marked changes in the rms-flux relation.  We find that one single component is required to explain the $\sim 40$ per cent variability observed at low count rates, whereas no or very low variability is associated to the accretion-disc thermal component.
\end{abstract}
\section{Introduction}
 
Black Hole X-ray transients (BHTs) are observed in different states during their outburst evolution. They correspond to particular spectral and timing properties seen in the observations. The physical interpretation of the states is still strongly discussed, but they are probably associated with different accretion regimes. A \textit{hard} state [historically known as low/hard state (LHS)] is generally observed at the beginning and at the end of the outburst. It is roughly characterized by a power-law shaped energy spectrum with a photon index of $\sim$ 1.6 (2-20 keV) and a high level of aperiodic variability [root mean square (rms) amplitude above $\sim$ 30 per cent]. In the middle of the outburst, the energy spectrum is dominated by a soft thermal component and almost no variability is seen. During this \textit{soft} state [historically known as high/soft state (HSS)]  a hard tail up to $\sim$ 1 MeV is also present (\citealt{Grove1998}). In contrast, a high-energy cut-off ( $\lesssim 200$ keV ) is observed in the LHS (e.g. \citealt{Wilms2006}; \citealt{Motta2009}). From a physical point of view, the thermal component present in the HSS is usually associated with emission from an optically thick accretion disc, whereas the emission in the hard state is believed to arise from a \textquoteleft corona\textquoteright~of hot electrons, where seed photons from an optically thin accretion disc are up-Comptonized. Synchrotron emission from a jet, which is known to dominate the radio and infrared spectrum during the LHS (e.g. \citealt{Fender2006}; \citealt{Russell2006}), could also have a significant contribution to the hard X-ray emission (\citealt{Russell2010}).\\
In between these two \textquoteleft canonical\textquoteright~states the situation is rather more complex. Hard-to-soft and a soft-to-hard transitions are observed in relatively short time scales (hours/days) as compared to the ones seen for the canonical states (weeks/months). During these transitions, both timing and spectral properties change dramatically, leading to different states classifications (see e.g. \citealt{Belloni2010} for a review). \cite{Homan2005} and \cite{Belloni2005} (B05) identify two additional states, the hard-intermediate state (HIMS) and the soft-intermediate state (SIMS) based on spectral and timing properties (see e.g \citealt{Casella2004} for different types of QPOs). \cite{McClintock2006} propose a more quantitative but model dependent classification, which requires of a steep power-law state and an intermediate state  in addition to the two canonical states (for a comparison see \citealt{Motta2009}; hereafter M09). \\  
A fundamental tool for studying the evolution of BHT is the hardness-intensity diagram (HID), where the evolution of the spectral hardness as a function of the acretion rate can be followed (see e.g. \citealt{Homan2001}). The hard--soft--hard usual evolution and the main transitions become apparent in the HID. The latter are usually not associated with major changes in flux, with the final soft-to-hard transition occurring at count-rates $\sim$ one order the magnitude lower. While the HID provides a general description of the BHT evolution, it is not enough for detailed studies (e.g. it is not able to establish with accuracy when the main transitions occur). Complementary tools like power density spectra (PDS), the hardness-rms correlation (e.g. B05), and multiwavelength observations are used for a more complete description of the different states and transitions. In this letter we present a new tool, the rms-intensity diagram, which allows one to map BHT states by only looking at the evolution of count-rate and rms (i.e. without spectral information). As a first step we have studied the case of the BHT GX 339-4, in which detailed observing campaigns have been performed during the multiple outburst observed with the \textit{Rossi X-ray timing explorer} (RXTE).\\
GX 339-4 is a low mass x-ray binary (LMXB) harbouring a $>6\sol$ accreting black hole (\citealt{hynes2003}; \citealt{tmd2008}). Since its discovery (\citealt{Markert1973}) the system has undergone several outburst, becoming one of the most studied X-ray transient. The intense monitoring carried out by RXTE during the 2002, 2004 and 2007 outburst has yielded detailed studies on the evolution of black hole states along the outburst (see e.g. B05). Here, we use this rich data set to study the evolution of the flux and variability along the outburst.
%The HID displayed by the source is very similar for the three cases (see e.g. \citealt{Belloni2010}) although the 2004 outburst is clearly fainter. %A new outburst is currently ongoing (\citealt{Yamaoka2010}) and numerous state transitions have been already observed with RXTE (\citealt{Motta2010a}; \citealt{Belloni2010a}; \citealt{Motta2010}).
\begin{figure*}
\centering
 \includegraphics[]{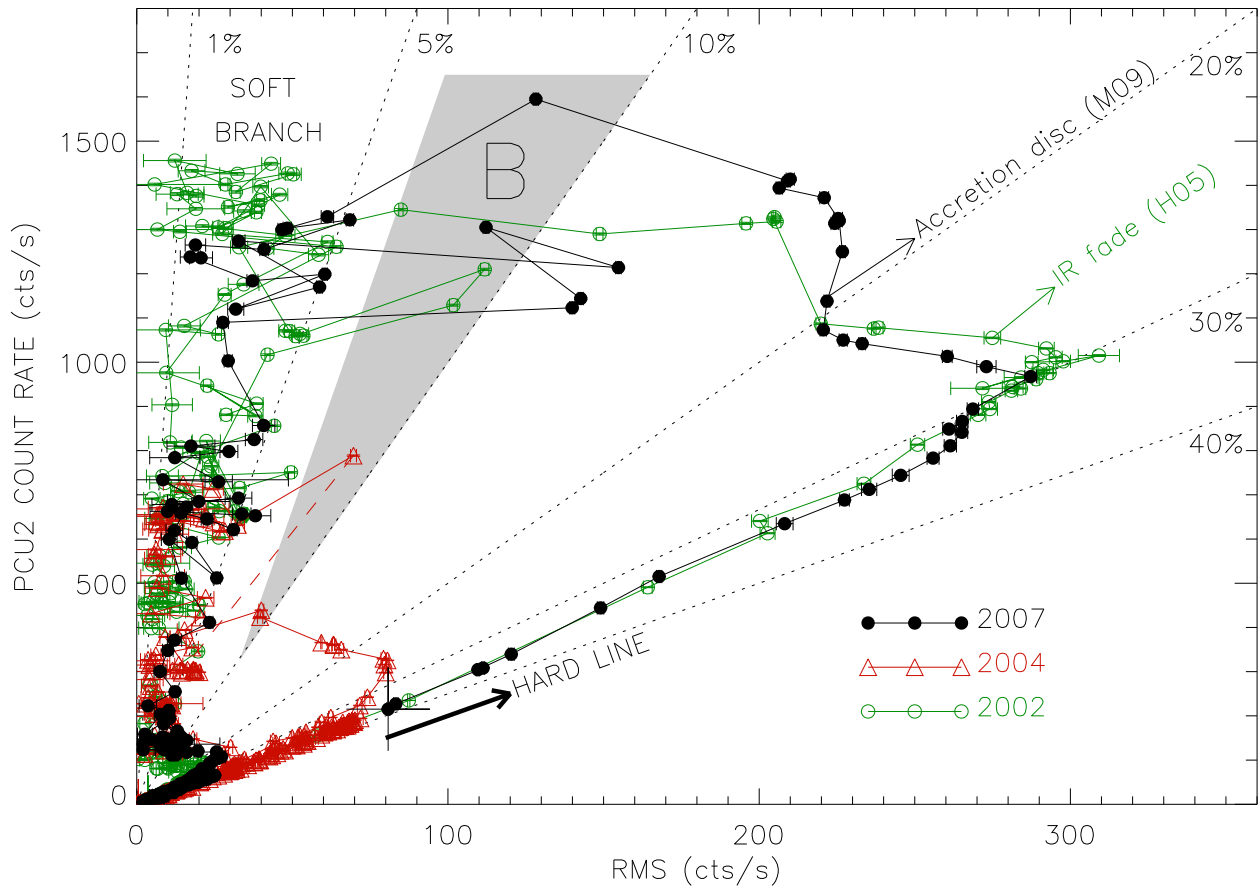}
\caption{Rms-intensity diagram for the 2002, 2004 and 2007 outbursts of GX 339-4. Dotted lines represent the 1, 5, 10, 20, 30 and 40 per cent fractional rms levels. The grey area marks the part of the diagram where solely type-B QPOs (\citealt{Casella2004}) are seen. The dashed line joins two consecutive observations separated by 33 days due to observability constraints. The cross corresponds to the first observation of the 2007 outburst.}  
\label{RID}
\end{figure*}

\section{Observations}
We have used all the RXTE public archival observations from the 2002 (206), 2004 (295) and 2007 (239) outburst of GX 339-4. For the analysis we have considered only the data from the \textit{Proportional Counter Array} (PCA). \textsc{Goodxenon}, \textsc{event} and \textsc{single-bit} data modes are used for the variability. Count-rates have been computed using PCA Standard 2 mode data corresponding to the PCU unit \#2. The analysis has been performed using \textsc{heasoft} V. 6.7. and custom timing software running under \textsc{idl}.\\  

\section{The rms-intensity diagram for the 2007 outburst}
The rms-intensity diagram (RID) for the 2007 outburst of GX 339-4 is presented in Fig.\ref{RID} as filled dots. For each observations net count rate corresponds to PCA channels 0--35 (2--15 keV). Only entire RXTE observations are initially considered, i.e. every dot corresponds to one observation. Power density spectra (PDS) for each observation have been also computed using the procedure outlined in \cite{Belloni2006}. We have used stretches 16 s long and the same energy selection as for the count rate. Total power was computed within the frequency band 0.1--64 Hz and the fractional rms calculated following \cite{Belloni1990}. The absolute rms displayed in the diagram (Fig.\ref{RID}) is obtained by multiplying fractional rms by net count-rate (PCU \#2) for each observation. The black solid line joins observations contiguous in time starting from the observation marked with a cross (i.e. observation \#1). We find that the source describe a continuous hysteresis pattern in the anticlockwise direction. Four different regions can be distinguished in this diagram.

\subsection{The hard line}      
\begin{figure}
\centering
 \includegraphics[width= 9cm,height=7cm]{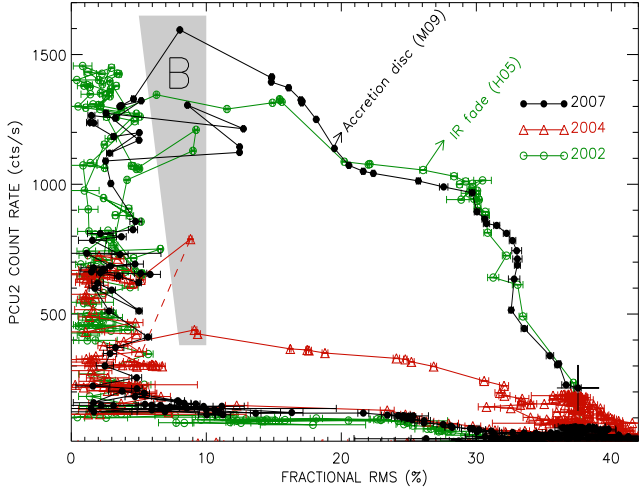}
\caption{Same as Fig.\ref{RID} but using fractional instead of absolute rms. Using this representation it is also possible to define a region where only type B QPOs are present.}
\label{RID_abs}
\end{figure}

Starting from the first observation (cross in Fig.\ref{RID}) and following the solid line we see rms increasing with count rate. %This linear relation observed at the base of the VD corresponds to the linear rms-flux relation originally observed by \cite{Uttley2001} in X-ray binaries and active galaxies. 
This linear trend is seen during the first 18 observations (i.e. $\sim 40$ days). Following the state classification reported by \cite{Motta2009}, all these observations belong to the canonical LHS. In the context of this digram we will call this linear relation  \textquoteleft hard line\textquoteright~ (HL). A continuous increase in hardness (M09) is observed during this HL.  Weak, type-C QPOs appear in the PDS of the observations corresponding to the upper part of the HL without modifying the rms-flux relation. They become strong once the system shows fractional rms of $\sim30$ per cent, and it is close to abandon the HL. In Fig.  \ref{RID_abs} we plot the RID but using fractional instead of absolute rms. In this representation the HL we see in Fig. \ref{RID} corresponds to the first 18 points starting from the cross, where fractional rms drops from 40 to 30 percent. 

\subsection{State transitions}
From observation \#19  to \#23 ($\sim 3$ days), the rms starts to decrease, whereas the flux is still increasing. Therefore, the system abandons the HL and follows a different linear relation. From observation \#24  to \#29 ($\sim 3$ days) rms a sudden increase in count rate is observed whereas rms is almost constant within $\sim $4 counts s$^{-1}$. The spectral fits performed by M09 require of a thermal black-body component from observation \#25 (see Fig. \ref{RID}). Thus, we associate this increase in flux at constant rms to the appearance (2-15 keV band) of an optically thick accretion disc with a very low variability level. \par
Until observation \#32, type-C QPOs are observed in the corresponding PDS. From observation \#32  to \#33 the rms decreases dramatically, crossing the 10 per cent fractional rms line. A type-B QPO is observed in the PDS. Although the change in rms is major, the observations are only separated by $\sim 1$ day.  

\subsection{The soft branch }
From observation \#34, the rms is below $\sim 5$ per cent and the count rate fades from $\sim$ 1300 to  $\sim$ 100 cts s$^{-1}$. A linear relation can be observed, but much more scattered than during the HL (see Fig. \ref{RID}). This new relation is followed by the system during the canonical soft state (M09).\\ 
A sudden increase in rms is also observed during four observations. Three of these cross the $\sim 10$ per cent line and show type-C QPOs whereas a type-B is observed for the other. Observations with type-A QPOs are not well differentiated from the soft branch. However, some of them are  placed close or slightly above the 5 per cent rms line, which roughly delimits the soft branch (see also Fig. \ref{RID_abs}).\\
\begin{figure}
\centering
 \includegraphics[width= 9cm,height=7cm]{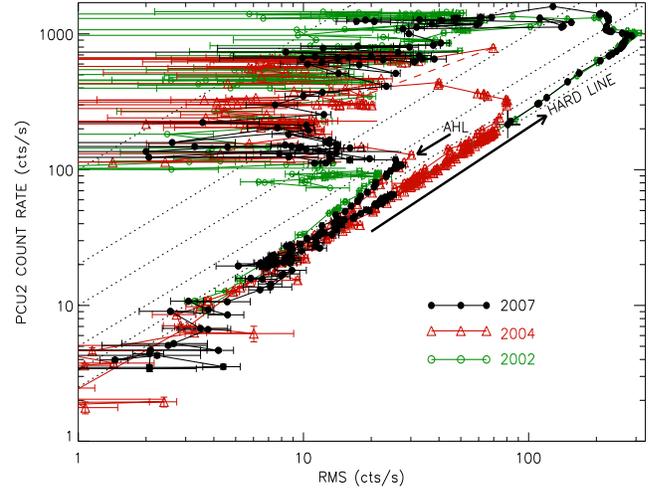}
\caption{Same as Fig.\ref{RID} but using a log-log scale. The  adjacent hard line (AHL) and the hard line at lower count rate becomes evident. The presence of a flux component with no variability is ruled out (see Sec. \ref{discussion}). Dotted lines represent the same fractional rms levels as in Fig.\ref{RID}.}
\label{RID_log}
\end{figure}
\subsection{Soft-to-hard transition. The adjacent hard line}
Fig. \ref{RID_log} shows the RID in log-log scale, where a monotonic rms increase is observed at constant flux ($\sim$ 100 cts s$^{-1}$) from the bottom of the soft branch. Once the rms cross the $\sim 20$ per cent level a linear decrease in rms and flux towards the HL is observed. During this adjacent HL (see note in Fig. \ref{RID_log} ) PDS are the typical of the LHS, although the spectral color is softer (i.e. hardness is lower) than the observed during the HL.

\section{The 2002 and 2004 outburst}
In Figs. \ref{RID}, \ref{RID_abs} and \ref{RID_log} the RID corresponding to the 2002 and 2004 outburst are shown as open triangles and open dots, respectively. 
\begin{itemize}
\item The 2002 RID is similar to the 2007 one (as in the case of the HID). The system leaves the HL roughly at the same position and an increase in flux at $\sim$ constant (absolute) rms is also seen in the region where the accretion disc is observed by M09 during 2007. \cite{Homan2005a} performed a multiwavelength campaign during the 2002 outburst. They find that the optical and near infrared  properties of GX 339-4 change on MJD 52398, which correspond to the first observation outside of the HL (see Fig. \ref{RID}). \cite{Belloni2005} also identify this observation as the first one belonging to the HIMS (i.e. non LHS). 
\item The 2004 ourburst is fainter and therefore its absolute rms is also significantly lower. In this case the variability does not decrease when the system leaves the HL, but increases with flux following a different linear relation. A clear feature related with the appearance of a thermal disc in the spectrum is not observed, suggesting a weaker disc component. Flux finally increases as rms decreases until the systems reach the soft branch. After a long permanence in this branch rms increases up to $\sim20$ per cent and goes back to the HL following the same adjacent HL observed in the other two outbursts.
\end{itemize}
\par
The HL is identical for the three outbursts, independent of the maximum flux reached. We also observe this behaviour in the BHT H1743-322 (Mu\~noz-Darias et al. in prep). The soft branch is also very similar between the three outbursts, lying within the $\sim1-5$ per cent fractional rms region. We notice that observations just outside this region contains fast state transitions (see \citealt{Nespoli2003}; \citealt{Casella2004}). A detailed state classification in this region of the diagram is beyond the scope of this work.\par
It is also possible to define a region  of the diagram (count rate $\gtrsim 400$ counts s$^{-1}$ and $\sim 7-10$ per cent fractional rms) where only observations showing type-B QPOs (grey region in Fig. \ref{RID} and Fig. \ref{RID_abs}) are seen. These observations  must  be classified as SIMS according to \cite{Homan2005}. We also identify two observations (70110-01-47-00 and 70108-03-02-00)  with fainter B QPOs at lower fractional rms than the region. A more detailed analysis of these data shows that the type-B QPO is only present  in part of the observation, with rms being the expected for the QPO B region.\\

\begin{figure}
\centering
 \includegraphics[width= 9cm,height=7cm]{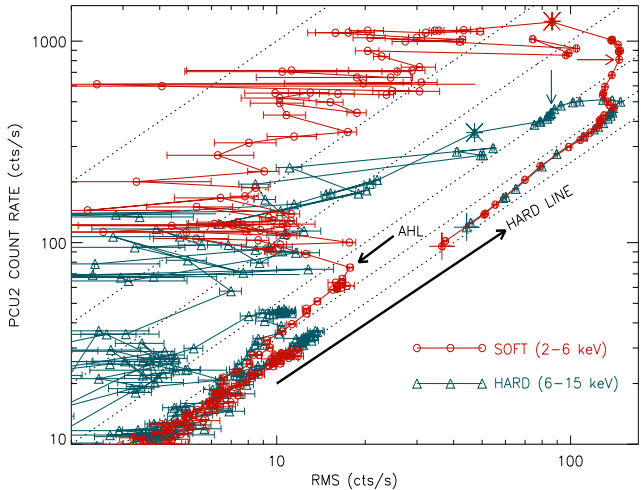}
\caption{Soft (PCA 0-13; 2-6 keV) and hard (PCA 14-35; 6-15 keV) rms-intensity diagrams for the 2007 outburst of GX 339-4 plotted in log-log scale. The dotted lines represent the same fractional rms levels as in Fig.\ref{RID}. Since an accretion disc component is detected in the energy spectrum (Motta et al. 2009; see arrows in the plot), the fractional rms level becomes higher in the hard band. the first observation showing a type-B QPO is indicated with a star. Crosses indicate the first observation.}
\label{RIDcl}
\end{figure}

\section{hard/soft rms-intensity diagrams}
With the aim of tracing the evolution of soft and hard variability we have computed the RID using soft (PCA 0-13; 2-6 keV) and hard (PCA 14-35; 6-15 keV) bands for the three outburst. These energy ranges can be selected independently of the observing mode in which data were taken, and allow us to cover the whole evolution of the three outbursts. In Fig. \ref{RIDcl} the soft and hard RID are plotted for the 2007 outburst as open circles and open triangles, respectively. It is clear that both RIDs show the same HL, where fractional rms drops from 40 to 30 percent rms as the system becomes brighter. This is expected since is well-known that rms is weakly energy dependent (2-20 keV) during the LHS (see e.g. \citealt{tmd2010} for the cases of XTE J1752-223 and Cyg X-1).
Differences arise when the system leaves the HL (observation \#18; 92035-01-02-04):
\begin{itemize}
\item During the next five observations the hard RID shows a decrease in rms at constant count-rate. Then, once the disc component appears in the spectrum (M09; see arrow in Fig. \ref{RIDcl}), a constant decrease in flux and rms following the 20 per cent line is seen until the type-B QPO appears (\#33; star in Fig. \ref{RIDcl}). For the same observations an increase of flux at constant rms is observed in the soft RID. 
\item Since the disc component is detected by M09 (see arrows in Fig. \ref{RIDcl}) , more fractional rms is seen systematically at higher energies, even at the very bottom of the soft branch.
\end{itemize}
On the other hand, the same adjacent HL is observed for both RID, as expected for LHS observations with flat rms spectra.\\ 
\subsection{The 2002 and 2004 outburst}
The behavior observed during the 2002 outburst (not shown here) is similar to the one described above. During 2004, the vertical rise after leaving the HL is much less evident in the soft RID, suggesting weaker accretion disc component. The decrease in rms at constant flux is neither observed in the hard RID, although the decay following the 20 per cent rms line is seen.

\section{Discussion} 
\label{discussion}  
During the outburst evolution of BHT different accretions regimes are present. They yield the different states that we see when looking at the X-ray emission arising from these systems. In each one of these states, one or more physical mechanisms (e.g. thermal or non-thermal Comptonization in the corona, synchrotron emission from a jet, thermal emission from an accretion disc) play a role. The relative contribution of these mechanisms can be studied using energy spectra, but also by analysing the particular variability they prompt in the light-curves. In this work we have shown that the evolution of the fast variability (0.1-64 Hz) can be used as a good tracer of the different accretion regimes in black hole binaries. This would be impossible in case the variability associated with the different spectral components were similar. Indeed, we find evidence for the presence of highly variable and an almost non variable components in the lightcurves of GX 339-4:
\begin{itemize} 
\item A sharp, linear relation between absolute rms and flux is observed at the beginning and at the end the three ourburst of GX 339-4 we have studied. This has been observed before by \cite{Gleissner2004} in Cyg X-1 during LHS observations. During these epochs, flux variations are probably tracing accretion rate variations and therefore rms is probably correlated with accretion rate. 
The rms-flux relation was firstly observed by \cite{Uttley2001} at much shorter time scales ($\sim 10$ s) than the ones presented here ($\sim 1$ ks). As a test we have studied the short term variations of the rms-flux relation by dividing the lightcurve in 32 s segments for the first and the last HL observations (2007 outburst). In both cases we also found a linear relation, with the fractional rms being consistent with the average of the entire observation. We note that a detailed study on these short-term variations is beyond the scope of this work.\\ 
\cite{Uttley2001} found that the extrapolation of the rms-flux relation was consistent with no variability at lower count rates than the ones they studied. This supports the presence of two different component in the lightcurves. GX 339-4 offers the opportunity to study this linear relation over a much broader flux range. Consistently with the found by \cite{Gleissner2004} by extrapolating the relation they obtain for Cyg X-1, we do not see  remaining flux when the rms is close to zero (see Fig. \ref{RID_log}), thus our results are consistent with the presence of one component solely responsible for the variability at low count rates. Indeed, the system reaches its maximum variability level when this component dominates the spectrum. However, in the upper part of the HL, where \cite{Uttley2001} perform their study, the fractional rms decreases from 40 to 30 per cent. This could be associated with the presence of a second, less variable component.
\item We find strong evidence for the presence of an extra component associated with the appearance of thermal emission in the energy spectrum of the source. This accretion disc component is not variable or varies very little. Its signature (i.e. flux increase at constant rms) is clearly seen in the 2002 and 2007 outburst once the systems leaves the HL. As expected, this becomes more evident if one only considers the 2-6 keV band (see Fig. \ref{RIDcl}). Indeed, The fractional variability of the system is lower at lower energies from the first observation in which the thermal component is detected until it returns to the LHS at the end of the outburst.  

\item  The behaviour of the hard (6-14 keV) RID is complex. Once the system leaves the HL, a decrease in rms at constant flux is observed, suggesting that the properties of the physical mechanism that drives the hard emission are different. This could be explained by invoking a change in the properties of the corona or the base of the jet. For instance, the high energy cut-off is observed to increase during the hard-to-soft transition (M09), which could be interpreted in terms of a cooling of the corona (see M09 and reference therein for details).  It is remarkable that very few observations show fractional rms $\lesssim 5$ per cent in the hard band, suggesting that an underlying variable component (probably associated to power-law emission) is present even during the soft state (see Fig. \ref{RIDcl}).\\  
\end{itemize}
\par
The different rms-flux relations outlined above enable to perform a state classification of the different observations using the rms-intensity diagram. The following states can be identified using the description by \cite{Homan2005}:
\begin{itemize}
\item Hard line observations corresponds to the LHS, with rms in the range $\sim$ 30--40 per cent.  During these observations PDS are qualitatively similar between each other, and can be fitted by using several broad Lorentzians (see e.g. Fig. 3 in \citealt{Belloni2005}, where the three upper PDS correspond to observations at the beginning, the middle and the end of the HL during the 2002 outburst). We associated the sharp changes we see in the rms-flux relation at the end of the HL to the transition to and from the HIMS. This is supported by spectral and timing changes (e.g. M09, B05) and particularly by the coincidence in time between the drop we see in rms and that observed in optical and infrared  by \cite{Homan2005a}.
\item During the HIMS, the rms is observed to drop from $\sim$ 30 to 10 per cent. The presence of the accretion disc in the spectrum is detected when the fractional rms is $\sim$ 20 per cent. We note that this point marks the transition between the hard and the steep power-law or intermediate states according to \cite{McClintock2006} (see M09). However, the major transition seems to occur before, when the system left the HL.
\item Soft-state observations are observed to follow a different rms-flux relation, consistent with the results obtained by \cite{Gleissner2004} using relatively soft observations of Cyg X-1. The soft state in GX 339-4 is much softer than in Cyg X-1 and a soft branch very different from the HL is clearly seen in the RID. It shows a variability level lower than $\sim 5$ per cent and much more scatter than the HL. This suggest fast changes in (at least) one of the spectral components of the lightcurve.  
\item We find that between $\sim 7-10$ per cent fractional rms only type-B QPOs are observed (grey regions in Fig. \ref{RID} and \ref{RID_abs}). They are SIMS observations according to \cite{Homan2005} and our diagram suggest that this kind of oscillations could follow a particular rms-flux relation. We note that transitions between soft and SIMS are very fast, which result in hybrid observations where timing properties are not constant. Thus, an accurate state classification close to the $\sim$ 5 per cent rms line is in some cases not compatible with using entire 1-3 ks observations.
\item An adjacent hard line is observed at the end of the three outbursts. Observations during this adjacent HL show PDS typical of the LHS but with a slighly softer spectrum than that of the HL. Why this other linear relation, not seen when the systems leaves the HL, is present is unclear. It could suggest the presence of new components (see \citealt{Russell2010}) or the total absence of any of the components (e.g. accretion disc) observed during the flux rise phase at the beginning of the outburst.       

\end{itemize}

\section{Conclusions}
We have proved that the evolution of the fast variability can be used as a good tracer of the different accretion regimes in black hole binaries. In particular, we find that apart from the linear rms-flux found during LHS, different relations are followed by black hole binaries during the soft and intermediate states.\\ 
In this work we have presented the rms-intensity diagram, showing that it is possible to associate the different regions of this diagram to the different states observed in black hole transients. Transitions can also be identified thanks to the marked changes they produce in the diagram. The application of this new tool to other systems and its interpretation will provide new insights in our understanding of the physical processes that take place during accretion episodes in X-ray binaries.
\vspace{0.5cm}

\noindent The research leading to these results has received funding from the European Community's Seventh Framework Programme (FP7/2007-2013) under grant agreement number ITN 215212 \textquotedblleft Black Hole Universe\textquotedblright. SM and TB acknowledge support from the ASI grant I/088/06/0 and PRIN-INAF 2008.

\bibliographystyle{mn2e.bst}
\bibliography{/Users/tmd/Documents/Liberia_RX.bib} 

\begin{thebibliography}{}

\bibitem[\protect\citeauthoryear{{Belloni} \& {Hasinger}}{{Belloni} \&
  {Hasinger}}{1990}]{Belloni1990}
{Belloni} T.,  {Hasinger} G.,  1990, \aap, 227, L33

\bibitem[\protect\citeauthoryear{{Belloni}, {Homan}, {Casella}, {van der Klis},
  {Nespoli}, {Lewin}, {Miller} \& {M{\'e}ndez}}{{Belloni}
  et~al.}{2005}]{Belloni2005}
{Belloni} T.,  {Homan} J.,  {Casella} P.,  {van der Klis} M.,  {Nespoli} E.,
  {Lewin} W.~H.~G.,  {Miller} J.~M.,    {M{\'e}ndez} M.,  2005, \aap, 440, 207

\bibitem[\protect\citeauthoryear{{Belloni}, {Parolin}, {Del Santo}, {Homan},
  {Casella}, {Fender}, {Lewin}, {M{\'e}ndez}, {Miller} \& {van der
  Klis}}{{Belloni} et~al.}{2006}]{Belloni2006}
{Belloni} T.,  {Parolin} I.,  {Del Santo} M.,  {Homan} J.,  {Casella} P.,
  {Fender} R.~P.,  {Lewin} W.~H.~G.,  {M{\'e}ndez} M.,  {Miller} J.~M.,    {van
  der Klis} M.,  2006, \mnras, 367, 1113

\bibitem[\protect\citeauthoryear{{Belloni}}{{Belloni}}{2010}]{Belloni2010}
{Belloni} T.~M.,  2010, Lecture Notes in Physics, Berlin Springer Verlag, 794,
  53

\bibitem[\protect\citeauthoryear{{Casella}, {Belloni}, {Homan} \&
  {Stella}}{{Casella} et~al.}{2004}]{Casella2004}
{Casella} P.,  {Belloni} T.,  {Homan} J.,    {Stella} L.,  2004, \aap, 426, 587

\bibitem[\protect\citeauthoryear{{Fender}}{{Fender}}{2006}]{Fender2006}
{Fender} R.,  2006, in Compact stellar X-ray sources, pp 381--419

\bibitem[\protect\citeauthoryear{{Gleissner}, {Wilms}, {Pottschmidt}, {Uttley},
  {Nowak} \& {Staubert}}{{Gleissner} et~al.}{2004}]{Gleissner2004}
{Gleissner} T.,  {Wilms} J.,  {Pottschmidt} K.,  {Uttley} P.,  {Nowak} M.~A.,
   {Staubert} R.,  2004, \aap, 414, 1091

\bibitem[\protect\citeauthoryear{{Grove}, {Johnson}, {Kroeger},
  {McNaron-Brown}, {Skibo} \& {Phlips}}{{Grove} et~al.}{1998}]{Grove1998}
{Grove} J.~E.,  {Johnson} W.~N.,  {Kroeger} R.~A.,  {McNaron-Brown} K.,
  {Skibo} J.~G.,    {Phlips} B.~F.,  1998, \apj, 500, 899

\bibitem[\protect\citeauthoryear{{Homan} \& {Belloni}}{{Homan} \&
  {Belloni}}{2005}]{Homan2005}
{Homan} J.,  {Belloni} T.,  2005, \apss, 300, 107

\bibitem[\protect\citeauthoryear{{Homan}, {Buxton}, {Markoff}, {Bailyn},
  {Nespoli} \& {Belloni}}{{Homan} et~al.}{2005}]{Homan2005a}
{Homan} J.,  {Buxton} M.,  {Markoff} S.,  {Bailyn} C.~D.,  {Nespoli} E.,
  {Belloni} T.,  2005, \apj, 624, 295

\bibitem[\protect\citeauthoryear{{Homan}, {Wijnands}, {van der Klis},
  {Belloni}, {van Paradijs}, {Klein-Wolt}, {Fender} \& {M\'endez}}{{Homan}
  et~al.}{2001}]{Homan2001}
{Homan} J.,  {Wijnands} R.,  {van der Klis} M.,  {Belloni} T.,  {van Paradijs}
  J.,  {Klein-Wolt} M.,  {Fender} R.,    {M\'endez} M.,  2001, \apjs, 132, 377

\bibitem[\protect\citeauthoryear{{Hynes}, {Steeghs}, {Casares}, {Charles} \&
  {O'Brien}}{{Hynes} et~al.}{2003}]{hynes2003}
{Hynes} R.~I.,  {Steeghs} D.,  {Casares} J.,  {Charles} P.~A.,    {O'Brien} K.,
   2003, \apjl, 583, L95

\bibitem[\protect\citeauthoryear{{Markert}, {Canizares}, {Clark}, {Lewin},
  {Schnopper} \& {Sprott}}{{Markert} et~al.}{1973}]{Markert1973}
{Markert} T.~H.,  {Canizares} C.~R.,  {Clark} G.~W.,  {Lewin} W.~H.~G.,
  {Schnopper} H.~W.,    {Sprott} G.~F.,  1973, \apjl, 184, L67+

\bibitem[\protect\citeauthoryear{{McClintock} \& {Remillard}}{{McClintock} \&
  {Remillard}}{2006}]{McClintock2006}
{McClintock} J.~E.,  {Remillard} R.~A.,  2006, pp 157--213

\bibitem[\protect\citeauthoryear{{Motta}, {Belloni} \& {Homan}}{{Motta}
  et~al.}{2009}]{Motta2009}
{Motta} S.,  {Belloni} T.,    {Homan} J.,  2009, \mnras, 400, 1603

\bibitem[\protect\citeauthoryear{{Mu{\~n}oz-Darias}, {Casares} \&
  {Mart{\'{\i}}nez-Pais}}{{Mu{\~n}oz-Darias} et~al.}{2008}]{tmd2008}
{Mu{\~n}oz-Darias} T.,  {Casares} J.,    {Mart{\'{\i}}nez-Pais} I.~G.,  2008,
  \mnras, 385, 2205

\bibitem[\protect\citeauthoryear{{Mu{\~n}oz-Darias}, {Motta}, {Pawar},
  {Belloni}, {Campana} \& {Bhattacharya}}{{Mu{\~n}oz-Darias}
  et~al.}{2010}]{tmd2010}
{Mu{\~n}oz-Darias} T.,  {Motta} S.,  {Pawar} D.,  {Belloni} T.~M.,  {Campana}
  S.,    {Bhattacharya} D.,  2010, \mnras, 404, L94

\bibitem[\protect\citeauthoryear{{Nespoli}, {Belloni}, {Homan}, {Miller},
  {Lewin}, {M{\'e}ndez} \& {van der Klis}}{{Nespoli}
  et~al.}{2003}]{Nespoli2003}
{Nespoli} E.,  {Belloni} T.,  {Homan} J.,  {Miller} J.~M.,  {Lewin} W.~H.~G.,
  {M{\'e}ndez} M.,    {van der Klis} M.,  2003, \aap, 412, 235

\bibitem[\protect\citeauthoryear{{Russell}, {Fender}, {Hynes}, {Brocksopp},
  {Homan}, {Jonker} \& {Buxton}}{{Russell} et~al.}{2006}]{Russell2006}
{Russell} D.~M.,  {Fender} R.~P.,  {Hynes} R.~I.,  {Brocksopp} C.,  {Homan} J.,
   {Jonker} P.~G.,    {Buxton} M.~M.,  2006, \mnras, 371, 1334

\bibitem[\protect\citeauthoryear{{Russell}, {Maitra}, {Dunn} \&
  {Markoff}}{{Russell} et~al.}{2010}]{Russell2010}
{Russell} D.~M.,  {Maitra} D.,  {Dunn} R.~J.~H.,    {Markoff} S.,  2010,
  \mnras, 405, 1759

\bibitem[\protect\citeauthoryear{{Uttley} \& {McHardy}}{{Uttley} \&
  {McHardy}}{2001}]{Uttley2001}
{Uttley} P.,  {McHardy} I.~M.,  2001, \mnras, 323, L26

\bibitem[\protect\citeauthoryear{{Wilms}, {Nowak}, {Pottschmidt}, {Pooley} \&
  {Fritz}}{{Wilms} et~al.}{2006}]{Wilms2006}
{Wilms} J.,  {Nowak} M.~A.,  {Pottschmidt} K.,  {Pooley} G.~G.,    {Fritz} S.,
  2006, \aap, 447, 245

\end{thebibliography}

\label{lastpage}
\end{document}